\documentclass[prd,twocolumn]{revtex4}
\usepackage{graphicx, epsfig}
\usepackage{color}
\usepackage{mathrsfs}
\usepackage{amsmath}

\newcommand{\be}{\begin{equation}}
\newcommand{\ee}{\end{equation}}
\newcommand{\bd}{\begin{equation*}}
\newcommand{\ed}{\end{equation*}}
\newcommand{\bea}{\begin{eqnarray}}
\newcommand{\eea}{\end{eqnarray}}

\newcommand{\gapp}{\mathrel{\raise.3ex\hbox{$>$}\mkern-14mu
              \lower0.6ex\hbox{$\sim$}}}
\newcommand{\lapp}{\mathrel{\raise.3ex\hbox{$<$}\mkern-14mu
              \lower0.6ex\hbox{$\sim$}}}

\begin{document}

\title{Classical and Quantum Equations of Motion for a BTZ Black String in AdS Space}
\author{Eric Greenwood}
\author{Evan Halstead}
\author{Peng Hao}
\affiliation{HEPCOS, Department of Physics,
SUNY at Buffalo, Buffalo, NY 14260-1500}
\begin{abstract}

We investigate gravitational collapse of a $(3+1)$-dimensional BTZ black string in AdS space in the context of both classical and quantum mechanics. This is done by first deriving the conserved mass per unit length of the cylindrically symmetric domain wall, which is taken as the classical Hamiltonian of the black string. In the quantum mechanical context, we take primary interest in the behavior of the collapse near the horizon and near the origin (classical singularity) from the point of view of an infalling observer. In the absence of radiation, quantum effects near the horizon do not change the classical conclusions for an infalling observer, meaning that the horizon is not an obstacle for him/her. The most interesting quantum mechanical effect comes in when investigating near the origin. First, quantum effects are able to remove the classical singularity at the origin, since the wave function is non-singular at the origin. Second, the Schr\"odinger equation describing the behavior near the origin displays non-local effects, which depend on the energy density of the domain wall. This is manifest in that derivatives of the wavefunction at one point are related to the value of the wavefunction at some other distant point.

\end{abstract}

\maketitle

\section{Introduction}

It is well know that gravity is the weakest of the four fundamental forces of nature, therefore it is expected that quantum effects will only become important in regions of very strong curvature, for example near classical singularities. These classical singularities are an endemic attribute of general relativity, as originally shown by Hawking and Penrose (see Ref.~\cite{HawkingPenrose}). However, it is expected that these singularities are only a reflection of our lack of an ultimate theory, one which is valid for all regimes. As a result, the general belief is that a full version of quantum gravity will rid gravitation of singularities, similar to how quantization rid other theories of similar singularity issues. 

A good theoretical backdrop to investigate gravitational quantization near classical singularities is in the context of black hole physics. A common example is found in the investigation of gravitational collapse of a massive object, i.e. a collapsing star or another gravitating body (see for example Ref.~\cite{Hawking}). In most cases, one considers a spherically symmetric object while taking into account different aspects of the body, such as mass, charge and angular momentum. However, an interesting question arises when one considers different topologies other than spherically symmetric. Does the topology of the collapsing object or background space-time play a role in the quantum effects of gravitational collapse?

To answer this question we consider the gravitational collapse of a BTZ black string in anti-De Sitter space (AdS). In 1992 it was discovered by Ba\~nados, Teitelboim, and Zanelli (see Ref.~\cite{btz}) that in $(2+1)$-dimensional gravity with negative cosmological constant, there are black hole solutions present. The existence of these objects, generally called BTZ black holes, is surprising given that the classical theory is ``trivial" in the absence of the cosmological constant, see for example Ref.~\cite{Witten}.  The solution was originally intended to be a lower-dimensional analog of general relativity, without the complications manifest in the higher-dimensional gravitational theory. However, the BTZ solution can be extended to the full $(3+1)$-dimensional theory by taking the direct product of the $(2+1)$-dimensional BTZ solution with another spatial dimension. It is well known that in $(3+1)$-dimensions, the BTZ solution takes the form of a cylinder, generally known as a black string. In this paper we wish to examine the classical and quantum solutions for gravitational collapse of the $(3+1)$-dimensional BTZ black string and investigate if they demonstrate similar properties found in spherically symmetric solutions, such as the Schwarzschild solution. Investigations of the classical equation of motion for the BTZ black string have been performed by several other authors in the past, see for example Refs.~\cite{MadhavGoswamiJoshi,DeshingkarJhinganChamorroJoshi,Eid}, while quantum solutions have been investigated in Refs.~\cite{PauloPitelliLetelier,Minassian}.

In Section \ref{sec:EQ} we review the Gauss-Codazzi equations. The Gauss-Codazzi equations investigate a three-dimensional hypersurface embedded in a four-dimensional space-time. In Section \ref{sec:BTZ} we derive the conserved mass per unit length for the BTZ black string using the Gauss-Codazzi equations found in Section \ref{sec:EQ}. In Section \ref{sec:classical} we study the classical equations of motion from the point of view of both an asymptotic and infalling (one who is falling together with the collapsing domain wall) observer. The most important results are presented in Section \ref{sec:Quantum} where we study the collapse from the point of view of an infalling observer in the context of quantum mechanics utilizing a minisuperspace version of the functional Schr\"odinger equation originally developed in Ref.~\cite{VachStojKrauss}. In Section \ref{sec:QM_Near} we explore quantum effects in the near-horizon limit for an infalling observer and show that, in the absence of quantum radiation, classical conclusions remain true, i.e. the horizon is no obstacle for an infalling observer. In Section \ref{sec:QM_0} we explore the quantum effects near the origin (i.e. classical singularity) and demonstrate two results: a) the wavefunction describing the collapsing domain wall is non-singular at the origin, and b) the non-local effects, which were absent at large distances, become unsuppressed in this near origin regime. 

We do remark that these conclusions are in the absence of quantum radiation, fields propagating in the background of the collapsing domain wall, which may introduce new elements. In Section \ref{sec:conclusion} we summarize our conclusions.

\section{The Equations}
\label{sec:EQ}

Here we follow the techniques developed in Ref.~\cite{Ipser}. Let $S$ denote a three-dimensional time-like hypersurface containing stress-energy and let $\xi^a$ be its unit spacelike normal ($\xi_a\xi^a=1$). The three-metric intrinsic to the hypersurface $S$ is 
\be
  h_{ab}=g_{ag}-\xi_a\xi_b
\ee
where $g_{ab}$ is the four-metric of the space-time. Here $h_{ab}$ is known as the projected tensor for the hypersurface $S$, see Ref.~\cite{Carroll}. This is due to the fact that, when acting $h_{ab}$ on a vector $v^a$, it will project it tangent to the hypersurface, hence orthogonal to $\xi^a$, 
\begin{align*}
  (h_{ab}v^a)\xi^b&=g_{ab}v^a\xi^b-\xi_a\xi_bv^a\xi^b\\
      &=v^a\xi_a-v^a\xi_a\\
       &=0.
\end{align*}
Let $\nabla_a$ denote the covariant derivative associated with $g_{ab}$ and let 
\be
  D_a=h_a{}^b\nabla_b,
  \label{D_a}
\ee
hence $D_a$ is the covariant derivative on the induced three-dimensional hypersurface. The extrinsic curvature of $S$, denoted by $\pi_{ab}$, is defind by
\be
  \pi_{ab}\equiv D_a\xi_b=\pi_{ba}.
  \label{pi}
\ee
The extrinsic curvature depends on how the hypersurface is embedded in the full four-dimensional space-time. The extrinsic curvature is used to differentiate different topologies. For example, intrinsic geometry of a cylinder and a torus can be flat, however, we know the exterior geometry of each is different. This different topology is given in the extrinsic curvature, which will tell us that we are actually on a torus or a cylinder. 

The contracted forms of the first and second Gauss-Codazzi equations are then given by
\bea
  ^3R+\pi_{ab}\pi^{ab}-\pi^2&=&-2G_{ab}\xi^a\xi^b\label{Gauss}\\
  h_{ab}D_c\pi^{ab}-D_a\pi&=&G_{bc}h^b{}_a\xi^c\label{Codazzi}.
\eea
Here $^3R$ is the Ricci scalar curvature of the three-geometry $h_{ab}$ of $S$, $\pi$ is the trace of the extrinsic curvature, and $G_a{}^b$ is the Einstein tensor in four-dimensional space-time.

Here we will be working with infinitely thin domain walls. The stress-energy tensor $T_{ab}$ of four-dimensional space-time then is assumed to have a $\delta$-function singularity on $S$. This in turn implies that the extrinsic curvature has a jump discontinuity across $S$, since the extrinsic curvature is analogous to the gradient of the Newtonian gravitational potential. Therefore we can introduce
\be
  \gamma_{ab}\equiv\pi_{+ab}-\pi_{-ab}
\ee
which is the difference between the exterior and interior extrinsic curvatures, and
\be
  S_{ab}\equiv\int dlT_{ab},
\ee
where $l$ is the proper distance through $S$ in the direction of the normal $\xi^a$, and where the subscripts $\pm$ refer to values just off the surface on the side determined by the direction of $\pm\xi^a$. Hence the direction for, say $+\xi^a$ will be in the direction of the exterior geometry of the domain wall, while $-\xi^a$ will denote the direction of the interior geometry of the domain wall. As we shall discuss below, these geometries will be different for the case of the spherically symmetric domain wall. Using Einstein's and the Gauss-Codazzi equations, one can show that (see Ref.~\cite{Wheeler})
\be
  S_{ab}=-\frac{1}{8\pi G_N}\left(\gamma_{ab}-h_{ab}\gamma_c{}^c\right).
  \label{action}
\ee
We can also introduce the ``average" extrinsic curvature
\be
  \tilde{\pi}_{ab}=\frac{1}{2}\left(\pi_{+ab}+\pi_{-ab}\right)
  \label{tilde_pi}
\ee
which will be important later.

\subsection{The Surface Stress-Energy Tensor}

Here we restrict ourselves to sources for which the stress energy tensor is given by, see Ref~.\cite{Ipser}
\be
  S^{ab}=\sigma u^au^b-\eta\left(h^{ab}+u^au^b\right)
  \label{Stress-Energy}
\ee
which is the material source consisting of a perfect fluid. In Eq.~(\ref{Stress-Energy}) $u^a$ is the four-velocity of any observer whose world line lies within $S$ and who sees no energy flux in his local frame, and where $\sigma$ is the energy per unit area and $\eta$ is the tension measured by the observer. For a dust wall it is well known that $\eta=0$, while for a domain wall $\eta=\sigma$. For a domain wall Eq.~(\ref{Stress-Energy}) reduces to 
\be
  S^{ab}=-\sigma h^{ab}.
\ee
We also note that the four-velocity $u^a$ is a time-like unit vector orthogonal to the space-like unit normal $\xi^a$, i.e.,
\begin{equation*}
  u_au^a=-1, \hspace{2mm} \xi_au^a=0, \hspace{2mm} \xi_a\xi^a=+1.
\end{equation*}

\subsection{Attractive Energy}

Here we derive equations for an observer who is hovering just above the surface $S$ on either side. Let the vector field $u^a$ be extended off $S$ in a smooth fashion. The acceleration
\bea
  u^a\nabla_au^b&=&(h^b{}_c+\xi^b\zeta_c)u^a\nabla_au^c\nonumber\\
     &=&h^b{}_cu^a\nabla_au^c-\xi^bu^au^c\pi_{ab}
\eea
has a jump discontinuity across $S$ since the extrinsic curvature has such a discontinuity. The perpendicular components of the accelerations of observers hovering just off $S$ on either side satisfy
\begin{align}
  \xi_bu^a\nabla_au^b\Big{|}_++\xi_bu^a\nabla_au^b\Big{|}_-=&-2u^au^b\tilde{\pi}_{ab}\nonumber\\
     =&-2\frac{\eta}{\sigma}(h^{ab}+u^au^b)\tilde{\pi}_{ab}
     \label{perp1}
\end{align}
and
\bea
  \xi_bu^a\nabla_au^b\Big{|}_+-\xi_bu^a\nabla_au^b\Big{|}_-&=&-u^au^b\gamma_{ab}\nonumber\\
     &=&4\pi G(\sigma-2\eta).
     \label{perp2}
\eea
Here we comment on the presence of the second term on the right hand side of Eq.~(\ref{perp1}). This term takes into account the contributions to the energy-tensor $T_{ab}$ which are present in the vacuo on opposite sides of $S$. For example, if there is only mass present, then $T_{ab}$ vanishes off the shell, hence the second term is zero. In the case of charge present, then $T_{ab}$ does not vanish, then the contribution to $T_{ab}$ outside can be taken from the Maxwell tensor. 

\section{BTZ Black String in (3+1) Dimensions}
\label{sec:BTZ}

For a cylindrical shell of stress-energy, let the unit normal $\xi_+$ point in the outward radial direction. It is well known that asymptotic flatness and cylindrical symmetry (assuming that the domain wall is infinitely long) requires that the interior geometry is flat (Birkhoff's theorem). For the external geometry we will choose an arbitrary metric. We will then consider the BTZ black string in $(3+1)$ dimensions.

Here we take the exterior metric as,
\begin{align}
  (ds^2)_+=&-f(r)dt^2+\frac{1}{f(r)}dr^2+r^2\left(d\phi^2+dz^2\right)
       \label{out_BTZ_metric}
\end{align}
and
\be
  (ds^2)_-=-A(r)dT^2+\frac{1}{A(r)}dr^2+r^2\left(d\phi^2+dz^2\right). \hspace{2mm} \text{for $r<R(t)$}
  \label{in_BTZ_metric}
\ee
Here the equation of the wall is 
\be
  r=R(t).
\ee
One finds for the components of $u^a$ and $\xi^a$ $(a=t$ or $T,r,\phi,z$, in that order$)$
\begin{align}
  (u^a{}_+)&=(\beta f^{-1},R_{\tau},0,0), \hspace{2mm} (u^a{}_-)=(\alpha A^{-1},R_{\tau},0,0),\nonumber\\
  (\xi^a{}_+)&=A(-R_{\tau},\beta f^{-1},0,0), \hspace{2mm} (\xi^a{}_-)=(-R_{\tau},\alpha A^{-1},0,0).
\end{align}
Here $R_{\tau}=dR/d\tau$, where $\tau$ is the proper time of an observer moving with four-velocity $u^a$ at the wall, and 
\begin{align}
  \alpha\equiv&AT_{\tau}=\sqrt{A+R_{\tau}^2},\label{alpha_BTZ}\\
  \beta\equiv&ft_{\tau}=\sqrt{f+R_{\tau}^2}.
  \label{beta_BTZ}
\end{align}
These expressions and the definitions Eqs.~(\ref{D_a}), (\ref{pi}) and (\ref{tilde_pi}) imply that
\be
  (h^{ab}+u^au^b)\tilde{\pi}_{ab}=(\xi^r{}_++\xi^r{}_-)\frac{1}{R},
\ee
and
\begin{align}
  \xi_bu^a\nabla_au^b\Big{|}_+&=\frac{1}{\beta}\left[R_{\tau\tau}+\frac{f'}{2}+\frac{(f+2R_{\tau}^2)R_{\tau}}{2f^2}\left(f_{\tau}-f'R_{\tau}\right)\right]\nonumber\\
  &=\frac{1}{\beta}\left[R_{\tau\tau}+\frac{f'}{2}\right]
\end{align}
\begin{align}
  \xi_bu^a\nabla_au^b\Big{|}_-&=\frac{1}{\alpha}\left[R_{\tau\tau}+\frac{A'}{2}+\frac{(A+2R_{\tau}^2)R_{\tau}}{2A^2}\left(A_{\tau}-A'R_{\tau}\right)\right]\nonumber\\
  &=\frac{1}{\alpha}\left[R_{\tau\tau}+\frac{A'}{2}\right]
  \label{acceleration}
\end{align}
where
\bd
  f'=\frac{df(r)}{dr}\Big{|}_{r=R(t)}
  \label{f_prime}
\ed
and
\bd
  A'=\frac{dA(r)}{dr}\Big{|}_{r=R(t)}.
  \label{A_prime}
\ed
In the second line in $a_+$ we used that 
\bd
  f_{\tau}=f'R_{\tau}
\ed
and in the second line in $a_-$ we used that 
\bd
  A_{\tau}=A'R_{\tau}.
\ed
Substituting into Eqs.~(\ref{perp1}) and (\ref{perp2}) then yields the equations of motion
\begin{align}
  (\alpha+\beta)R_{\tau\tau}=&-2\frac{\eta}{\sigma}\frac{\alpha\beta(\alpha+\beta)}{R}-\frac{\alpha f'}{2}-\frac{\beta A'}{2}\label{plus}\\
  (\alpha-\beta)R_{\tau\tau}=&4\pi\alpha\beta G(\sigma-2\eta)-\frac{\alpha f'}{2}+\frac{\beta A'}{2}.\label{minus}
\end{align}
Taking the ratio of Eqs.~(\ref{plus}) and (\ref{minus}) allows us to eliminate $R_{\tau\tau}$ from the expression.

Here we make some general comments on Eqs.~(\ref{plus}) and (\ref{minus}). $R_{\tau\tau}$ is always negative provided $\eta,f',A'\geq0$. Hence a cylindrical domain wall with $\eta\geq0$ will always collapse to a black hole, regardless of its size. A qualitatively similar result was found in Ref.~\cite{Olea}, where the authors consider a Hamiltonian treatment of the gravitational collapse.

From Ref.~\cite{ChaoGuang} we take that the metric coefficient $f$ is given as
\be
  f=-\frac{\Lambda}{3}R^2-\frac{4GM}{R}
  \label{f}
\ee
and
\be
  A=-\frac{\Lambda}{3}R^2
  \label{A}
\ee
where $M$ is the mass per unit length in the $z$ direction. From the ratio of Eqs.~(\ref{plus}) and (\ref{minus}) we then find the mass to be given by
\begin{align}
  M&=\pi\sigma R^2\left(\alpha+\beta\right)\nonumber\\
   &=\pi\sigma R^2\left(\sqrt{-\frac{\Lambda}{3}R^2+R_{\tau}^2}+\sqrt{-\frac{\Lambda}{3}R^2-\frac{4GM}{R}+R_{\tau}^2}\right).
   \label{btz_mass}
\end{align}
Solving explicitly for the mass one finds
\be
  M=2\pi\sigma R^2\left[\sqrt{-\frac{\Lambda}{3}R^2+R_{\tau}^2}-2\pi\sigma GR\right].
  \label{BTZ_mass}
\ee
It can be checked that Eq.~(\ref{BTZ_mass}) is a constant of motion, which is done in Appendix \ref{ch:check}.

From Eqs.~(\ref{f}) and (\ref{A}) we see that Eq.~(\ref{plus}) is always negative, i.e. collapsing, provided that $\Lambda\leq0$.

\section{Classical Equations of Motion}
\label{sec:classical}

Here we investigate the classical equations of motion. As discussed in Section \ref{sec:BTZ}, the shell only collapses when $\Lambda<0$, thus hereafter we will only consider negative cosmological constant. 

The effective action consistent with the conserved quantity, Eq.~(\ref{BTZ_mass}), is given by
\begin{align}
  S_{eff}=-2\pi\sigma\int d\tau R^2&\Big{[}\sqrt{\frac{\Lambda}{3}}R\sqrt{1+3\frac{R_{\tau}^2}{\Lambda R^2}}\nonumber\\
     &-R_{\tau}\sinh^{-1}\left(\sqrt{\frac{3}{\Lambda R^2}}R_{\tau}\right)-2\pi\sigma GR\Big{]}.
   \label{IN_action}
\end{align}
To proceed we wish to examine these equations from two different viewpoints, those of the asymptotic and infalling observers, respectively. First we will consider the asymptotic observer, followed by that of the infalling observer. 

\subsubsection{Asymptotic Observer}

Here we are interested in the viewpoint of the asymptotic observer. Therefore we will transform Eq.~(\ref{IN_action}) to the viewpoint of the asymptotic observer with the help of Eq.~(\ref{beta_BTZ}) as
\begin{align}
  S_{eff}=-2\pi\sigma\int dt R^2\Big{[}&\sqrt{\frac{\Lambda R^2}{3}}\sqrt{f-\frac{(1-3(\Lambda R^2)^{-1}f)}{f}\dot{R}^2}\nonumber\\
     &-\dot{R}\sqrt{f}\sinh^{-1}\left(\sqrt{\frac{3}{\Lambda R^2}}\dot{R}\sqrt{\frac{f}{f^2-\dot{R}^2}}\right)\nonumber\\
     &-2\pi\sigma GR\sqrt{f-\frac{\dot{R}^2}{f}}\Big{]}
  \label{AS_action}
\end{align}
Therefore the effective Lagrangian expressed in terms of the asymptotic observer's time $t$ is given by
\begin{align}
  L_{eff}=-2\pi\sigma R^2\Big{[}&\sqrt{\frac{\Lambda R^2}{3}}\sqrt{f-\frac{(1-3(\Lambda R^2)^{-1}f)}{f}\dot{R}^2}\nonumber\\
     &-\dot{R}\sqrt{f}\sinh^{-1}\left(\sqrt{\frac{3}{\Lambda R^2}}\dot{R}\sqrt{\frac{f}{f^2-\dot{R}^2}}\right)\nonumber\\
     &-2\pi\sigma GR\sqrt{f-\frac{\dot{R}^2}{f}}\Big{]}
  \label{AS_L}
\end{align}

The generalized momentum, $\Pi$, can be derived from Eq.~(\ref{AS_L})
\begin{align}
  \Pi=\frac{2\pi\sigma R^2}{\sqrt{f}}\Big{[}&\frac{3}{\Lambda R^2}\frac{f^3\dot{R}}{(f^2-\dot{R}^2)\sqrt{f^2-(1-9(\Lambda R^2)^{-2}f)\dot{R}^2}}\nonumber\\
    &+\sqrt{\frac{\Lambda R^2}{3}}\frac{\dot{R}(1-3(\Lambda R^2)^{-1}f)}{\sqrt{f^2-(1-3(\Lambda R^2)^{-1}f)\dot{R}^2}}\nonumber\\
    &+f\sinh^{-1}\left[\dot{R}\sqrt{\frac{3}{\Lambda R^2}}\sqrt{\frac{f}{f^2-\dot{R}^2}}\right]\nonumber\\
    &-\frac{2\pi\sigma GR\dot{R}}{\sqrt{f^2-\dot{R}^2}}\Big{]}
  \label{AS_Pi}
\end{align}
Thus the Hamiltonian in terms of $\dot{R}$ is given by
\begin{align}
  H=&\frac{2\pi\sigma R^2}{\sqrt{f}}\Big{[}\frac{3}{\Lambda R^2}\frac{f^3\dot{R}^2}{(f^2-\dot{R}^2)\sqrt{f^2-(1-9(\Lambda R^2)^{-2}f)\dot{R}^2}}\nonumber\\
    &+\sqrt{\frac{\Lambda R^2}{3}}\frac{f^2}{\sqrt{f}\sqrt{f^2-(1-3(\Lambda R^2)^{-1}f)\dot{R}^2}}-\frac{2\pi\sigma GRf^2}{\sqrt{f}\sqrt{f^2-\dot{R}^2}}\Big{]}.
    \label{AS_H}
\end{align}

To obtain $H$ as a function of $(R,\Pi)$, we need to eliminate $\dot{R}$ in favor of $\Pi$ using Eq.~(\ref{AS_Pi}). This can, in principle, be done but is very combersome. Instead we consider the limit when $R$ is close to $R_H$ and $f\rightarrow0$. In the limit $f\rightarrow0$ the denominators in Eq.~(\ref{AS_Pi}) are equal, therefore we can write
\be
  \Pi\approx\frac{2\pi\mu R^2\dot{R}}{\sqrt{f}\sqrt{f^2-\dot{R}^2}}
  \label{AS_apPi}
\ee 
where
\be
  \mu\equiv\sqrt{\frac{\Lambda R_H^2}{3}}-2\pi\sigma GR_H
\ee
and where $R_H$ is the horizon radius. Using Eq.~(\ref{AS_H}) we can then write the Hamiltonian as
\be
  H\approx\frac{2\pi\mu f^{3/2}R^2}{\sqrt{f^2-\dot{R}^2}}=\sqrt{(f\Pi)^2+f(2\pi\mu R^2)^2}.
  \label{AS_apH}
\ee
Here we note that Eq.~(\ref{AS_apH}) has the form of a relativistic particle, $\sqrt{p^2+m^2}$, with a position-dependent mass term. 

The Hamiltonian is a conserved quantity and so, from Eq.~(\ref{AS_apH}),
\be
  h=\frac{f^{3/2}R^2}{\sqrt{f^2-\dot{R}^2}}
  \label{AS_h}
\ee
where $h=H/2\pi\mu$ is a constant. Solving Eq.~(\ref{AS_h}) for $\dot{R}$ we get
\be
  \dot{R}=\pm f\sqrt{1-\frac{fR^4}{h^2}},
\ee
which, in the near horizon limit, takes the form
\be
  \dot{R}\approx\pm f\left(1-\frac{1}{2}\frac{fR^4}{h^2}\right)
\ee
since $f\rightarrow0$ as $R\rightarrow R_H$.

The dynamics for $R\sim R_H$ can be obtained by solving the equation $\dot{R}=\pm f$. However, as $R\sim R_H$ we can write
\bd
  f\approx\frac{\Lambda R_H^2}{3}-\frac{4GM}{R}
\ed 
this then gives, to leading order in $R-R_H$,
\be
  R\approx R_H+(R_0-R_H)e^{-\Lambda t/3R_H}
  \label{R_t}
\ee
where $R_0$ is the radius of the domain wall at $t=0$. As we are interested in the collapsing solution, we choose the negative sign in the exponent. This solution implies that, from the classical point of view, the asymptotic observer never sees the formation of the horizon of the black hole, since $R(t)=R_H$ only as $t\rightarrow\infty$. Thus, the time needed for a collapsing object to cross its own horizon radius is infinite from a point of view of a static outside observer.

\subsubsection{Infalling Observer}

The effective action for the infalling observer is given in Eq.~(\ref{IN_action}). Therefore the effective Lagrangian expressed in terms of the infalling observer's time $\tau$ is given by
\begin{align}
  L_{eff}=&-2\pi\sigma R^2\Big{[}\sqrt{\frac{\Lambda}{3}}R\sqrt{1+3\frac{R_{\tau}^2}{\Lambda R^2}}\nonumber\\
     &-R_{\tau}\sinh^{-1}\left(\sqrt{\frac{3}{\Lambda R^2}}R_{\tau}\right)-2\pi\sigma GR\Big{]}.
   \label{IN_L}
\end{align}

The generalized momentum, $\tilde{\Pi}$, can be derived from Eq.~(\ref{IN_L})
\be
  \tilde{\Pi}=2\pi\sigma R^2\sinh^{-1}\left(\sqrt{\frac{3}{\Lambda R^2}}R_{\tau}\right).
  \label{IN_Pi}
\ee
The Hamiltonian (in terms of $R_{\tau}$) is 
\be
  H=2\pi\sigma R^2\left[\sqrt{\frac{\Lambda R^2}{3}+R_{\tau}^2}-2\pi\sigma GR\right]
  \label{IN_H}
\ee
which is just Eq.~(\ref{BTZ_mass}). 

From Eq.~(\ref{IN_H}) we can calculate $R_{\tau}$ to be
\be
  R_{\tau}=\pm\sqrt{\left(\frac{\tilde{h}}{R^2}+2\pi\sigma GR\right)^2-\frac{\Lambda R^2}{3}}
  \label{R_tau}
\ee
where $\tilde{h}=H/(2\pi\sigma)$. In our formalism, the Hamiltonian is just the conserved mass, $H=M$ (the value of which can be viewed as an initial condition). As a zeroth order approximation of Eq.~(\ref{R_tau}), we can see that in the regime $R\sim R_H$ the velocity is a constant
\bd
  R_{\tau}\approx\pm\sqrt{\left(\frac{\tilde{h}}{R_H^2}+2\pi\sigma GR_H\right)^2-\frac{\Lambda R_H^2}{3}}
\ed
with solutions
\be
  R=R_0-\tau\sqrt{\left(\frac{\tilde{h}}{R_H^2}+2\pi\sigma GR_H\right)^2-\frac{\Lambda R_H^2}{3}}
  \label{R_IN_0}
\ee
where we have chosen the minus sign since the shell is collapsing and $R_0$ is the initial position of the domain wall at $\tau=0$. As another approximation, we take the limit that $\Lambda$ is small. The solution to Eq.~(\ref{R_tau}) is then
\be
  R=\left(\frac{1}{2\pi\sigma G}\left[\left(h+2\pi\sigma GR_0^3\right)e^{-6\pi\sigma G\tau}-h\right]\right)^{1/3}.
  \label{R_IN_1}
\ee

\begin{figure}[hptb]
\includegraphics{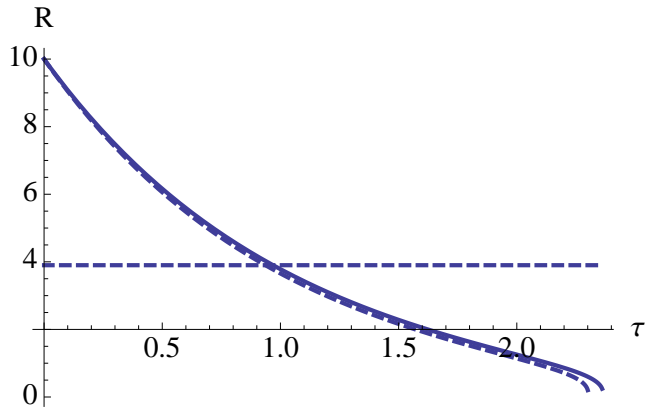}
\caption{Here we plot $R$ versus $\tau$ for solution of Eq.~(\ref{R_tau}), represented by the solid line, and the approximation in Eq.~(\ref{R_IN_1}), represented by the dashed curve. Here the dashed horizontal line is the position of the horizon.}
\label{R_tau_plot}
\end{figure}

By inspection, we can see that both Eqs.~(\ref{R_IN_0}) and (\ref{R_IN_1}) give that the domain wall will cross the horizon, with respect to the infalling observer, in a finite amount of proper time. In Figure \ref{R_tau_plot} we compare the numerical solution of Eq.~(\ref{R_tau}) versus the approximation given in Eq.~(\ref{R_IN_1}). As can be seen, even in the full numerical solution the domain wall crosses the horizon within a finite amount of proper time. We can also see that the domain wall will collapse to the classical singularity in a finite amount of propertime as well.

\section{Quantum Mechanical Collapse}
\label{sec:Quantum}

\subsection{Asymptotic Observer}

In this section, we will quantize the Hamiltonian in terms of the asymptotic observer time $t$ defined in Eq.~(\ref{AS_apH}).

The classical Hamiltonian in Eq.~(\ref{AS_apH}) has a square root and so we first consider the squared Hamiltonian
\be
  H^2=f\Pi f\Pi+f(2\pi\mu R^2)^2
  \label{H2}
\ee
where we have made a choice for ordering $f$ and $\Pi$ in the first term. Now we apply the standard quantization procedure. We substitute
\be
  \Pi=-i\frac{\partial}{\partial R}
\ee
into the squared Schr\"odinger equation,
\be
  H^2\Psi=-\frac{\partial^2\Psi}{\partial t^2}.
\ee
Then we have
\be
  -f\frac{\partial}{\partial R}\left(f\frac{\partial\Psi}{\partial R}\right)+f(2\pi\mu R^2)^2\Psi=-\frac{\partial^2\Psi}{\partial t^2}.
  \label{H2_Schrod_R}
\ee
To solve this equation, we define
\be
  u=\int \frac{dR}{f}
  \label{u}
\ee
which gives
\be
  f\Pi=-i\frac{\partial}{\partial u}.
\ee
Eq.~(\ref{H2_Schrod_R}) then gives
\be
  \frac{\partial^2\Psi}{\partial t^2}-\frac{\partial^2\Psi}{\partial u^2}+f(2\pi\mu R^2)^2=0.
  \label{H2_Schrod_u}
\ee
This is just the massive wave equation in a Minkowski background with a mass term that depends on position. Note that $R$ needs to be written in terms of the coordinate $u$ and this can be done, in principle, by inverting Eq.~(\ref{u}). However, care needs to be taken to choose the correct branch since the region $R\in(R_H,\infty)$ maps onto $u\in(-\infty,\infty)$ and $R\in(0,R_H)$ maps onto $u\in(0,-\infty)$. 

We are interested in the situation of a collapsing domain wall. In the region $R\sim R_H$, the logarithm in Eq.~(\ref{u}) dominates. We look for wave-packet solutions propagating toward $R_H$, toward $u\rightarrow-\infty$. In this limit
\be
  f\sim e^{u}\rightarrow0
\ee
and the last term in Eq.~(\ref{H2_Schrod_u}) can be ignored. Wave-packet dynamics in this region is simply given by the free wave equation, and any function of light-cone coordinates $(u\pm t)$ is a solution. In particular, we can write a Gaussian wave-packet solution that is propagating toward the horizon 
\be
  \Psi=\frac{1}{\sqrt{2\pi}z}e^{-(u+t)^2/2z^2}
\ee
where $z$ is some chosen width of the wave-packet in the $u$ coordinate. The width of the Gaussian wave-packet remains fixed in the $u$ coordinate, while it shrinks in the $R$ coordinate via $dR=fdu$ which follows from Eq.~(\ref{u}). However, to get to the horizon, it must travel out to $u=-\infty$, and this takes an infinite amount of asymptotic observer time. So we can conclude that the quantum domain wall does not collapse to $R_H$ in a finite time, as far as the asymptotic observer is concerned, so that quantum effects considered here do not alter the classical result that an asymptotic observer does not observe the formation of an event horizon.  

\subsection{Infalling Observer}

In this section, we will quantize the Hamiltonian in terms of the infalling observer time $\tau$ defined in Eq.~(\ref{BTZ_mass}), near the horizon and near the classical singularity.

\subsubsection{Near horizon}
\label{sec:QM_Near}

The exact Hamiltonian is given by Eq.~(\ref{BTZ_mass}). However, here again we have the difficulty of the square root term. Unlike in the previous section, simply considering the Hamiltonian squared will not get rid of the square root. Therefore we have to consider the regular Hamiltonian. To simplify the analysis, we will consider the near horizon limit as well as the limit where $R_{\tau}$ is small. This is indeed a restriction to special motion of the domain wall, since in general $R_{\tau}$ can be large near $R_H$ if the domain wall is collapsing from a very large distance. However, one may always choose initial conditions in such a way that the initial position of the domain wall $R(\tau=0)$ is very close to $R_H$.

In the limit of small $R_{\tau}$ and $R\sim R_H$ the Hamiltonian simplifies to 
\be
  H=2\pi\sigma R_H^2\left[\sqrt{\frac{\Lambda R^2_H}{3}}\left(1+\frac{3}{2\Lambda R_H^2}R_{\tau}^2\right)-(2\pi\sigma GR_H)\right].
\ee
In the same limit, the momentum reduces to 
\be
  \tilde{\Pi}=2\sqrt{\frac{3}{\Lambda}}\pi\sigma R_HR_{\tau}.
  \label{tilde_Pi_n0}
\ee
Dropping the constant terms from the Hamiltonian we get
\be
  H=\sqrt{\frac{\Lambda}{3}}\frac{\tilde{\Pi}^2}{2\pi\sigma R_H}.
  \label{In_Ham_RH}
\ee
Using the standard quantization procedure, we substitute
\be
  \tilde{\Pi}=-i\frac{\partial}{\partial R}
\ee
into Eq.~(\ref{In_Ham_RH}). This then gives us
\be
  -\sqrt{\frac{\Lambda}{3}}\frac{1}{2\pi\sigma R_H}\frac{\partial^2\Psi}{\partial R^2}=i\frac{\partial\Psi}{\partial\tau}.
\ee
This is just the Schr\"odinger equation for a freely propagating ``particle" of mass $2\pi\sigma R_H\sqrt{3}/\sqrt{\Lambda}$, as we expected in this approximation. Since $R_H$ is only a finite distance away for an infalling observer we can conclude that quantum effects do not alter the classical result that a collapsing domain wall crosses its own horizon in a finite amount of proper time. 

\subsubsection{Near the Classical Singularity}
\label{sec:QM_0}

In this section we investigate the most important question of quantum effects for the infalling observer, when the collapsing domain wall approaches the origin (i.e. the classical singularity at $R\rightarrow0$). The exact value for $R_{\tau}$ is given in Eq.~(\ref{R_tau}). Near the origin, i.e. in the limit $R\rightarrow0$ (keeping only the leading order terms) becomes
\be
  R_{\tau}\approx-\frac{\tilde{h}}{R^2}
\ee
where again $\tilde{h}=H/(2\pi\sigma)$, while the Hamiltonian is just the conserved mass $H=M$. This implies that, up to the leading term near the origin, the Hamiltonian is 
\be
  H=2\pi\sigma R^2R_{\tau}.
  \label{H_tau_n0}
\ee
Substituting the asymptotic behavior in Eq.~(\ref{H_tau_n0}) in the expression for the generalized momentum given in Eq.~(\ref{IN_Pi}), we learn that 
\be
  \lim_{R\rightarrow0}\tilde{\Pi}=0
\ee
and
\be
  \lim_{R\rightarrow0}\frac{\tilde{\Pi}}{2\pi\sigma R^2}=-\infty.
\ee
This implies that $R_{\tau}$ defined as
\be
  R_{\tau}=\sqrt{\frac{\Lambda R^2}{3}}\sinh\left(\frac{\tilde{\Pi}}{2\pi\sigma R^2}\right)
\ee
near the origin becomes
\be
  R_{\tau}=\frac{1}{2}\sqrt{\frac{\Lambda R^2}{3}}\exp\left(-\frac{\tilde{\Pi}}{2\pi\sigma R^2}\right)
  \label{R_tau_n0}
\ee
Therefore we can write the Schr\"odinger equation as
\be
  \sqrt{\frac{\Lambda}{3}}\pi\sigma R^3\exp\left(\frac{i}{2\pi\sigma R^2}\frac{\partial}{\partial R}\right)\Psi=i\frac{\partial\Psi}{\partial\tau}.
  \label{IN_Schrod_n0}
\ee

In Eq.~(\ref{IN_Schrod_n0}), the differential operator in the exponent gives some unusual properties
to the equation. Note that if we expand the exponent we can not stop the series after
a finite number of terms but need to include all of the terms. This means that we need
to include an infinite number of derivatives of the wave function $\Psi$ into the differential
equation. An infinite number of derivatives of a certain function uniquely specifies the
whole function. Thus, the value of (the derivative of) the function on the right hand side
of Eq.~(\ref{IN_Schrod_n0}) at one point depends on the values of the function at different points on the left hand side of the same equation. This is in strong contrast with ordinary local differential
equations where the value of the function and certain finite number of its derivatives are
related at the same point of space. This indicates that Eq.~(\ref{IN_Schrod_n0}) describes physics which is not strictly local.

With a simple change of variables we can make this argument more transparent. If
we introduce a new variable $v = R^3$, Eq.~(\ref{IN_Schrod_n0}) becomes
\be
  \sqrt{\frac{\Lambda}{3}}\pi\sigma v\exp\left(\frac{3i}{2\pi\sigma}\frac{\partial}{\partial v}\right)\Psi=i\frac{\partial\Psi(v,\tau)}{\partial\tau}.
\ee
The differential operator in the exponent is just a translation operator which shifts the
argument of the wave function by a non-infinitesimal amount of $3i/(2\pi\sigma)$. Since the wave
function is complex in general, a shift by a complex value is not a problem. Therefore
Eq.~(\ref{IN_Schrod_n0}) can be written as
\be
  \sqrt{\frac{\Lambda}{3}}\pi\sigma v\Psi\left(v+\frac{3i}{2\pi\sigma},\tau\right)=i\frac{\partial\Psi(v,\tau)}{\partial\tau}
  \label{IN_Schrod_v}
\ee

The wave function near the origin $\Psi(R\rightarrow0,\tau)$ is in fact related to the wave function
at some distant point $\Psi(R\rightarrow(\frac{3i}{2\pi\sigma})^{1/3},\tau)$. At large distances far from the origin, non-local effects were absent (at least in the approximation we used) as can be seen from Eq.~(\ref{tilde_Pi_n0}) which gives a linear relation between the generalized velocity and the generalized momentum. However, in the last stages of the collapse, when $R\rightarrow0$, these effects become important. Eq.~(\ref{R_tau_n0}) contains the generalized momentum (and thus the derivatives) in the exponent which makes the Schr\"odinger equation non-local. From Eq.~(\ref{IN_Schrod_v}) we see that non-locality depends on the wall surface tension $\sigma$. For light walls (small $\sigma$) non-locality is stronger, while large $\sigma$ suppresses it. However, the suppression can not be arbitrarily large since $\sigma$ cannot be too large either. Otherwise one would start off inside of the Horizon.

While it is possible that the whole formalism breaks down at very short distances of the order of Planck length, it is obvious that there will be a regime in which non-local effects are important.

Eq.~(\ref{IN_Schrod_v}) also implies that the wave function describing the collapsing shell is nonsingular at the origin. Indeed, in the limit of $R\rightarrow0$, this equation becomes
\be
  \frac{\partial\Psi(R\rightarrow0,\tau)}{\partial\tau}=0
  \label{Non_Singular}
\ee
where we used the fact that the wave function at some finite $R$, i.e. $\Psi(R\rightarrow(\frac{3i}{2\pi\sigma})^{1/3},\tau)$, is finite. From Eq.~(\ref{Non_Singular}) it then follows that $\Psi(R\rightarrow0)=\text{const}$. Non-singularity of the wave function describing the collapsing object at the origin is very important knowing that the origin represents the classical singularity and it is the source of most of the problems and paradoxes in black hole physics. Some consequences will be discussed in conclusions.

\section{Conclusion}
\label{sec:conclusion}

We studied gravitational collapse of a cylindrically symmetric BTZ black string (in (3+1)-dimensions) represented by a thin cylindrically symmetric domain wall with negative cosmological constant in the context of quantum mechanics. To do so, we first found the conserved mass per unit length using the the Gauss-Codazzi formalism. Due to the structure of the conserved mass, we identified it as the Hamiltonian of the system. The Hamiltonian allowed us to solve both the classical and quantum equations of the collapsing domain wall. To find the quantum solution, we employed the inherently quantum functional Schrodinger formalism, which can be readily incorporated into the Wheeler-de Witt formalism. A general treatment of the full Wheeler-de Witt equation is very difficult, thus we employed the minisuperspace version by truncating the field degrees of freedom to a finite subset. We examined most of the cases of interest.

In the literature, there exist some arguments that quantum effects near the horizon might change what an asymptotic observer would see in gravitational collapse. In particular, it was argued that quantum fluctuations can make the collapse time finite for a static outside observer. However, it appears that, at least in the framework of the functional Schr\"odinger formalism, quantum effects do not change the conclusions of classical general relativity, i.e.~it takes infinite time according to the asymptotic observerÕs clock for the collapsing domain wall to cross its own horizon, similar to the result found for the spherically symmetric domain wall found in Ref.~\cite{GreenStoj}. This conclusion is only valid in the absence of quantum radiation. Hawking radiation of course makes the lifetime of the black hole finite and asymptotic observers will see infalling matter accumulated onto the stretched horizon (see for example Refs.~\cite{PriceThorne,SusskindThorlaciusUglum}), thermalized, and eventually re-emitted as part of the Hawking radiation.

Results from the point of view of an infalling observer (an observer who is falling together with the collapsing domain wall) are of special interest. We first investigated what an infalling observer would see when the domain wall is crossing its own Schwarzschild radius. To do that, we explored quantum effects in near horizon limit for an infalling observer and showed that, in the absence of quantum radiation, classical conclusions remain true, i.e. horizon is no obstacle for an infalling observer. This result is again the same as for the spherically symmetric case studied in Ref.~\cite{GreenStoj}.

Finally, we explored quantum effects near the origin (i.e. classical singularity) from the point of view of an infalling observer. There are two important effects to be mentioned. First, the wave function describing the collapsing shell is non-singular at the origin. This result is in agreement with those found in Ref.~\cite{PauloPitelliLetelier} (for a massless test particle), Ref.~\cite{GreenStoj} and Ref.~\cite{WangGreenStoj} (when considering a Reisner-Nordstr\"om domain wall), and contrary to those found in Ref.~\cite{Minassian}, where the author used a quantum generalized affine parameter. The finding of the non-singular wave function at the origin is also in agreement with the expectation that quantization will rid gravity of singularities, just as in atomic physics it got rid of the singularity of the Coulomb potential which has an identical classical $1/r$ behavior. If the singularity at the origin is really erased, then most of the assumed properties of the black holes need to be re-thought. In particular, once we include quantum radiation, a collapsing object (or a black hole) will lose all of its energy in finite time. However, in the absence of the true singularity at the center, a horizon formed during the collapse can not be a true global event horizon. In other words, in the
absence of the singularity, a Òblack holeÓ may trap the light and other particles for some finite amount of time, but not forever (nor is the information lost down the singularity). This has profound implications for black hole physics.

Second, the quantum equation that governs physics near the classical singularity seems to be non-local, as first seen in Refs.~\cite{GreenStoj,WangGreenStoj} in the context of the functional Schr\"odinger equation. The Schr\"odinger equation describing the collapsing object contains an infinite number of derivatives. The dynamics of the wave function at a certain point near the origin depends on the value of the wave function at some distant point. While these nonlocal effects were absent at large distances far from the origin, they become unsuppressed in the near origin regime. Non-local effects we are finding in our approach may signal two things. It may be that it is a simple consequence of the fact that the Wheeler-de Witt (or functional Schrodinger) formalism is only an approximation of some more fundamental local theory. The other possibility is that the quantum description of the black hole physics requires inherently non-local physics. The answer to this question requires further investigation.

\section*{Acknowledgments}

The authors are grateful to D. Stojkovic, for very useful discussions. 

\appendix

\section{Check that Mass per unit Length is a constant of Motion}
\label{ch:check}

Here we wish to check that Eq.~(\ref{BTZ_mass}) is a constant of motion. To do so, we start with Eq.~(\ref{btz_mass}). Taking the derivative of Eq.~(\ref{btz_mass}) with respect to $\tau$ gives
\begin{align}
  M_{\tau}=&\pi\sigma R\Big{[}2R_{\tau}(\alpha+\beta)\nonumber\\
    &+R\left(\frac{1}{2\alpha}(A_{\tau}+2R_{\tau}R_{\tau\tau})+\frac{1}{2\beta}(f_{\tau}+2R_{\tau}R_{\tau\tau})\right)\Big{]}
\end{align}
where we used Eqs.~(\ref{alpha_BTZ}) and (\ref{beta_BTZ}). Now from the structure of the metric coefficients, we can use that 
\bd
  A_{\tau}=R_{\tau}A',
\ed
and
\bd
  f_{\tau}=R_{\tau}f'.
\ed
Hence we can write the derivative of the mass per unit length with respect to $\tau$ as
\begin{align}
  M_{\tau}=&\pi\sigma R\Big{[}2R_{\tau}(\alpha+\beta)+RR_{\tau}R_{\tau\tau}\left(\frac{1}{\alpha}+\frac{1}{\beta}\right)\nonumber\\
    &+\frac{RR_{\tau}}{2}\left(\frac{f'}{\beta}+\frac{A'}{\alpha}\right)\Big{]}.
\end{align}
Using Eq.~(\ref{plus}) for a domain wall we arrive at
\be
  M_{\tau}=0.
\ee
Hence the mass per unit length is a conserved quantity.

\end{document}